# Concept of Feedback in Future Computing Models to Cloud Systems

Evgeniy V. Pluzhnik, Evgeniy V. Nikulchev, Simon V. Payain

Moscow Technological Institute «WTU», Moscow, Russia
e.pluzhnik@gmail.com, nikulchev@mail.ru, sadsema@gmail.com

**Abstract.** Currently, it is urgent to ensure QoS in distributed computing systems. This became especially important to the development and spread of cloud services. Big data structures become heavily distributed. Necessary to consider the communication channels and data transmission systems and virtualization and scalability in future design of computational models in problems of designing cloud systems, evaluating the effectiveness of the algorithms, the assessment of economic performance data centers. Requires not only the monitoring of data flows and computing resources, but also the operational management of these resources to QoS provide. Such a tool may be just the introduction of feedback in computational models. The article presents the main dynamic model with feedback as a basis for a new model of distributed computing processes. The research results are presented here. Formulated in this work can be used for other complex tasks - estimation of structural complexity of distributed databases, evaluation of dynamic characteristics of systems operating in the hybrid cloud, etc.

## 1    Introduction

Cloud technologies and services are now developing. However, with the increasing prevalence of clouds is the growth number outstanding task. Under cloud computing, as a rule, understand Internet services provided by specialized data centers in the form of hardware and software system, or a distributed computing system consisting of a set of interconnected virtual machines, allowing dynamic compute resources to provide a certain level of service [1].

Despite the new technology in cloud computing include long-established principles of virtualization, routing, code duplication, etc. In general, it uses the known computational model. New device structure networks require the development of new guidelines for the future of computing processes. It seems obvious innovations such as use feedbacks. In computing systems are long gone only consecutive and parallel computing. There are queues, waiting for service, the ability to distribute computing resources.

Another problem put hybrid cloud infrastructure. When designing systems not clear to us how to partition information without performance degradation. There is a question about the effectiveness of existing transmission systems and databases in the cloud. Obviously, to obtain the gain, it is necessary to develop criteria for the system transmission efficiency. A drawback system operating in the clouds is attached to communication channels. However, the use of feedback provides the opportunity to move to the dynamic control of the model. Criteria developed for them and the corresponding optimization techniques. Using control systems to evaluate the structural complexity, as have the ability to oversee all operational units, including passage through the channels and the current state of resources. We conducted an experiment to confirm this assertion. [2] Object of study used semi-structured data to a large database. In the scientific and educational environment, as a rule, are widely used in systems of this data - this articles, tutorials, quizzes, etc. As a result, found that when using the cloud we have not lost in productivity even a single request. A query using threads was achieved higher quality of service.

Publications in the field of query optimization in cloud data centers are conducted in the last two or three years. It may be noted studies [3, 4, 5, 6]. In connection with the optimization of queries there are quite a number of problems: problems of query transformation to a more effective non-procedural representation (logic optimization), the problem of choosing a set of alternative procedural query execution plans , problems of cost estimates for the selected query execution plan , etc. The problems associated with the logical query optimization, has created a direction called semantic optimization. So many researchers are problems valuations procedural query execution plans.

To optimize queries traditionally used model graph theory, algorithm theory and other methods of discrete mathematics. They cannot get the correct assessment of conformity of theoretical research into practical implementations in the cloud technologies. This is largely due to the fundamentally different building information systems - it is not only the distributed data warehouse , but also the use of virtualization with dynamic reallocation of resources, the use of communication channels with different bandwidth for query processing, various platform features. All this has led to the search for new tools description of database management systems.

State space representation in the form of a system of equations is not very known in the art to optimize the database and therefore not covered by these experts. The article therefore has a slightly unusual structure. It begins with an example which cannot be described with the help of traditional formal models and optimization tools. And only then provides a brief overview of the successful use of dynamic models for cloud resource management.

It is proposed to use to evaluate the performance of systems operating in hybrid environments, to use the terminology and methods of system analysis. That certainly is not new, at the levels of service PaaS [1] cloud infrastructure is widely used concept of automatic control, automatic allocation of resources using the methods of dynamics.

For computing systems are intuitive terminologies of structural complexity, observability, reachability of these information systems [7, 8]. That means all of the many well-developed tools and control theory can be applied to study the parameters of information systems.

Given the logic of the cloud computing is difficult to determine the effectiveness of the algorithms. Virtualization and scalability of resources may unexpectedly greatly speed up the algorithms, as in our experimental example. That's a probably will help identify methods of system identification.

In general, it can be noted that the transfer system is necessary to develop a stable structure, the study of the characteristics of reliability, observability. Decomposition of the local and cloud components should be based on the methods of structural complexity of systems analysis.

The dynamic descriptions in the form of approximation of differential equations with the vector control were used in the different tasks that are close to the considered in the article. In [9] the processes that include real-time embedded systems. You can also note the earlier article for managing web servers [10], and virtualized data centers [11].

In previous works, such as [12], an approach that assumes that the cloud can be modeled as a Multi-Input-Multi-Output (MIMO), the system is implemented for capacity control in the cloud. However, the design of the regulars and identification systems do not allow the full use of these ideas. In the work [13] proposed to use Single-Input-Single-Output (SISO) systems with the introduction of the notion proportional thresholding.

The original paper [14] is devoted to the construction of a dynamic model of linked servers in the cloud. The important results of this work are the introduction of the concept of positive feedback, the theoretical proof of the stability of the model, the use of passive systems.

The closest in terms of models of our article [15]. This article describes how to identify patterns in the problem space and formulated the principles of the use of automatic control cloud resources.

Control theory was used in telecommunication systems recently. Our team of dynamic models has been used for solving problems of modeling of network traffic (this model is given as an example in the article [16]).

Dynamic issues are considered and applied to ensure QoS. In [17] discusses the use of a guaranteed rate transmission mechanism for maintaining QoS during network congestion. In [18] the development of a routing protocol to ensure QoS using temporary bandwidth reservation. In [19] deals with the modeling of traffic for the mobile network based on QoS.

## 2   Structure of Hybrid Cloud

Consider the standard hybrid cloud structure (Fig. 1). There shows a basic map of virtualization technology used in cloud infrastructure. There are four levels of allocated. At the first level two zones - local physical server, server virtualization. Second and third level is the switching and routing. The fourth level is the public cloud and the global network.

Private cloud - level of hardware virtualization with support for special processor architecture, with the possibility of reallocation of computing resources allocated between VM.

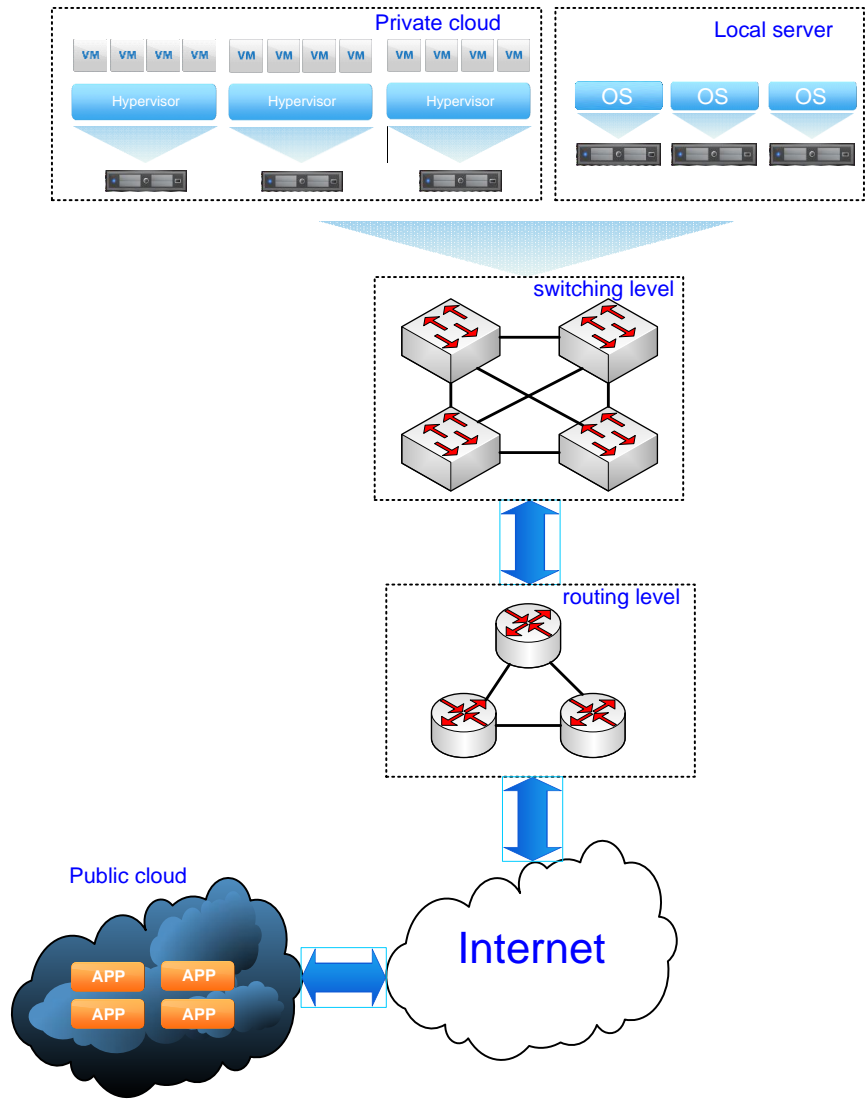

**Fig. 1. Structure of hybrid cloud**

Exchange of data between virtual machines occurs at the level of programmatic data network inside the level to control. Local server level is a set of servers without of virtualization, where each server has a single operating system. These servers are processing and storage. These servers are processing and storage.

Exchange of data between servers occurs at the level of switching and routing. At the level Routing level occurs packet routing, setting their priorities to control the queues, service levels and capacity allocation between the transmitted data types.

At the level of global Internet data is transmitted is not guaranteeing the delivery of the level of service, as well as the route of the packet.

At the level of Public cloud computing is an allocation of the necessary resources for processing and storage of information without the possibility of changes in their allocation and distribution.

Transition through levels requires not only monitor but also control. Precisely in these transitions and data loss occurs. Currently, the problem is solved multiple overlapping data.

Must be admitted that solve these problems, developers of industrial technology intuitively system. Of course, with intuitive approach may be situations resonance, when all the time of computing resources will be employed in the analysis process, and not present the solution of problems. We kind of watched it in the first multitasking operating systems for personal computers. It is therefore necessary to develop a basic theoretical position and conduct research on the equipment in order to create new computational models focused on cloud technologies.

## 3    Dynamic Models: Basic Assumptions

Let us consider basic positions which presence constitute grounds for an approach to the control system theory and dynamic models.

Consider a distributed database in the cloud. In cloud databases exist feature is that there are software or hardware routing. They store data on how the virtual machine on which pieces of data are stored, how much is in the public, private cloud which determines the level of security and other service information. In fact, even the same type of incoming query is converted in each case to a different scenario and route maps and depends on the model structure, and communication channels.

Input and output parameters for the study are the parameters characterizing the cloud computing system resources - CPU usage is at a given time, channel load, the control signals on the state of the virtual machines and clusters. In some cases, you can consider the input of a query that can be measured, for example, in the perfect disjunctive normal form.

An important assumption! The system cannot be considered only at a single query, there is a constant stream of requests. The system can be retrieved, then released. Therefore the state of the system at time $t+1$ depends on the query, and the current state at time $t$. In addition, there are natural constraints imposed by the width of channels, the number of free processors, the amount of free memory, etc.

As you know, in control theory it is transient response of the system to a single signal. A single signal is usually scaled signal outputting system to normal operation, i.e. changes settings when turned on. It is easy to get in a situation where the "bad structure of the clouds" will be oscillatory processes. It will be in a situation where the growth will stimulate queries rotation system increases and decreases in resources. Perhaps monotonous output to normal. All of this means that you must modeling and solution of the classical problems for which was established control theory - stability analysis in the design of feedback systems, the synthesis of optimal controllers, control software.

The presence of feedback, first defined by the need to consider the current state of the virtual machine workload. As was done in [20]. Secondly for distributed databases final data are delivered to the client via the cloud environment, and information about the end of the query to be delivered to the central system. It is understood that the presence of more feedbacks charged already narrow communication channels and may increase the processing time. This is one more argument in favor of the control systems for the optimization of feedback.

Feature of the cloud is positive feedback.

The base model offers classic model is approximated by the following difference equations,

$$x(t+1) = Ax(t) + Bu(t),$$

where $x = (x_1, x_2, ..., x_n)^T$ – n-dimensional vector of the system states under given constraints $x \in X \subseteq R^n$, $u = (u_1, u_2, ..., u_m)^T$ – m-dimensional vector of controls under given constraints $u \in U \subseteq R^m$, $t$ – discrete time instant.

In previous articles on the use of the control CPU, control of power in the data center and cloud computing [13, 21], a linear stationary system implementation found to be adequate.

It should be admitted, however, that in some cases and in the case of non-linear systems their identification is justified [16, 22].

It is natural that when you transfer to the cloud structure of the databases and route data retrieval becomes difficult. In the simple example given in the first part of the beginning of article [2] was introduced by an additional block that separates the search request in the private and public cloud. So even at the top level representation of cloud databases becomes difficult to follow without mathematical methods for the structural stability of the system [15]. Therefore connected to the structure of the system requires the solution of tasks, known for the theory of management - sustainability assessment framework, assessment of structural complexity, etc.

## 4       Experiments

Developed method of the required parameters monitoring and control. Management scheme is shown in Figure 2.

We obtain a result to ensure the QoS by dynamically traffic prioritization at routing.

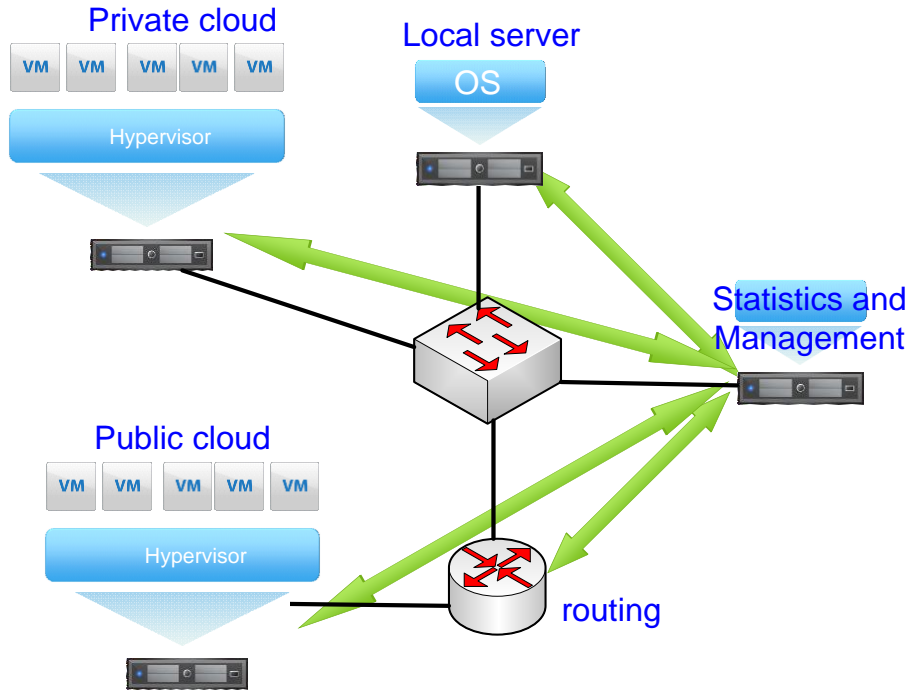

**Fig. 2. Management scheme**

Enlarged algorithm consists of the following steps.
1. Allocation of traffic types.
2. Monitoring traffic types.
3. If none of the channel for any type of traffic is approaching 70%, dynamic control module is included.
4. If the predicted value (for the forecast horizon) senior priority traffic increases, an increase in the width of the channel for a given type on the predicted value.
5. If the predicted value (for the forecast horizon) senior priority traffic decreases, the observed decrease in the channel width for a given type on the predicted value.
6. If there is an increase in traffic i and i +1 priority, then a decrease in traffic on the smaller priority value prediction, but no more than the critical value.

Based on data on the Internet download channel data obtained during the monitoring of the corporate network for each month, measured throughout the year, was built empirical histogram frequency channel load (Fig. 3a). Statistics obtained by removing information from the router interfaces on the amount of data transferred and loading port.

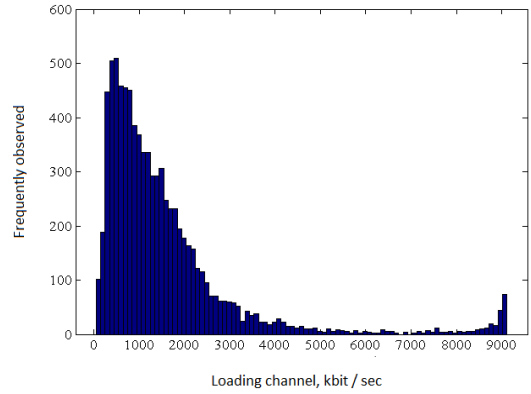

a)

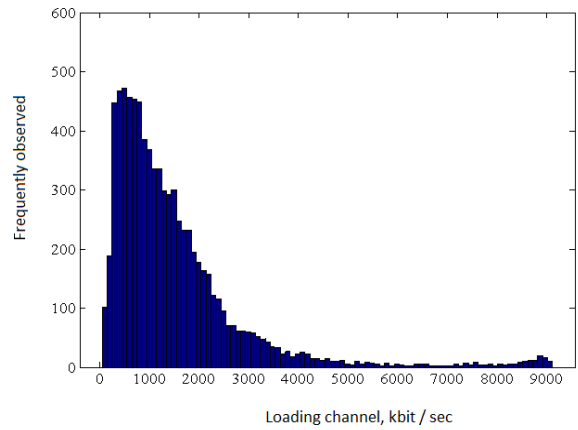

b)

**Fig. 3. Results of studies: a) without control, b) with control.**

As a result of the control algorithm developed for the picture has changed, the histogram shown in Fig. 4b. It is seen that lost their tails in the distribution and peaks. This means that the number of dropped packets dropped and increased quality of data delivery.

Note that the control can be used not only predictive model. Perhaps provide management and control by the reference model. It is harvested when several models and at some point of time is monitored indicators and the next time step is used one or another of prefabricated model.

## 5      Conclusion

The present work is aimed at attracting the attention of specialists in the field of cloud database to the apparatus control theory. Presented in this article allows the use of the description of the known methods for solving important practical problems. In conclusion, allow to formulate them.

Investigation of the stability of systems - the most important solution for the presence of feedback in distributed systems with disabilities through the communication channel, limited by safe access to the data, as well as scalable resources under peak load.

The study of controllability - for complex structured systems, the problem of controllability in a given time interval, the task of reachability and observability is not obvious without the pilot study, the use of the proposed model allows the use of famous mathematical tool solutions.

Designing effective systems - for large databases with complex levels of access and distributed in different systems with different hardware and software support, as it is observed in a hybrid cloud infrastructure.

**REFERENCE**

1. Buyya, R., Calheiros, R.N., Xiaorong Li. Autonomic Cloud computing: Open challenges and architectural element // 3th Int. Con. Emerging Applications of Information Technology (EAIT), pp. 3–10. (2012) (doi: 10.1109/EAIT.2012.6407847)
2. Pluzhnik E.V., Nikulchev E.V. Use of dynamical systems modeling to hybrid cloud database // International Journal of Communications, Network and System Sciences,  vol. 6. no. 12, p. 505-512. (2013)  (doi: 10.4236/ijcns.2013.612054)
3. Silva Y. N., Larson P., Zhou J. Exploiting common subexpressions for cloud query processing // Data Engineering (ICDE), 2012 IEEE 28th International Conference, p. 1337–1348. (2012) (doi: 10.1109/ICDE.2012.106)
4. Jurczyk P., Xiong L. Dynamic query processing for p2p data services in the cloud // Database and Expert Systems Applications, p. 396-411. Springer, Berlin (2009) (doi: 10.1007/978-3-642-03573-9_34)
5. Wang, X., Liu, X., Fan, L., Jia, X. A Decentralized Virtual Machine Migration Approach of Data Centers for Cloud Computing  // Mathematical Problems in Engineering, vol. 2013, article ID 878542,10pages . (2013) (doi: 10.1155/2013/878542)
6. Sithole E., McConnell A., McClean S., Parr G., Scotney B., Moore A., Bustard D. Cache performance models for quality of service compliance in storage clouds // Journal of Cloud Computing: Advances, Systems and Applications, vol. 2, no. 1, pp. 1–24. (2013) (doi: 10.1186/2192-113X-2-1)
7. Zhendong Sun, S.S. Ge Analysis and synthesis of switched linear control systems // Automatica, vol. 41,  no. 2, pp. 181–195 (2005) (doi: 10.1016/j.automatica.2004.09.015).
8. Ohta, Y.,  Maeda, H., Kodama, S. Reachability, Observability, and Realizability of Continuous-Time Positive Systems Read // SIAM Journal on Control and Optimization, vol. 22, no. 2, (1984) (doi: 10.1137/0322013)

9. Lu C., Wang X., Koutsoukos X. Feedback utilization control in distributed real-time systems with end-to-end tasks // Parallel and Distributed Systems, IEEE Transactions. vol. 16, no. 6, pp. 550-561. (2005) (doi: 10.1109/TPDS.2005.73)
10. Diao, Y., Gandhi, N., Hellerstein, J. L., Parekh, S., Tilbury, D. M. Using MIMO feedback control to enforce policies for interrelated metrics with application to the Apache web server // Network Operations and Management Symposium, 2002. pp. 219-234. IEEE (2002). (doi: 10.1109/NOMS.2002.1015566)
11. Xu W., Zhu, X., Singhal, S., & Wang, Z. Predictive control for dynamic resource allocation in enterprise data centers // Network Operations and Management Symposium, 2006. 10th IEEE/IFIP. – IEEE, pp. 115-126. (2006) (doi: 10.1109/NOMS.2006.1687544)
12. Nandina, V., Luna, J. M., Nava, E. J., Lamb, C. C., Heileman, G. L., & Abdallah, C. T. Policy-based Security Provisioning and Performance Control in the Cloud // Int. Conf. on Cloud Computing and Services Science (CLOSER 2013). SciTePress, pp. 502-508. (2013)
13. Nathuji, R., Kansal, A., Ghaffarkhah, A. Q-clouds: Managing performance interference effects for QoS-aware clouds // Proc. the ACM European Society in Systems Conference. pp. 237–250. Paris, France (2010) (doi: 10.1145/1755913.1755938)
14. Lim, H., Babu, S., Chase, J., Parekh, S. Automated control in cloud computing: Challenges and opportunities // Proc. 1st Workshop on Autom. Ctrl for Datacenters & Clouds, pp. 13–18. Barcelona. (2009) (doi: 10.1145/1555271.1555275)
15. Lemmon, M. D. Towards a passivity framework for power control and response time management in cloud computing //  Proc. of 7th Intl. Workshop on Feedback Computing,  San Jose, CA. (2012)
16. Nikulchev, E., Kozlov, O. Identification of Structural Model for Chaotic Systems // Journal of  Modern Physics, vol. 4, no. 10, pp. 1381-1392. (2013) (doi: 10.4236/jmp.2013.410166).
17. Yu, B., Xu, C. Efficient QoS Scheme in Network Congestion // Lecture Notes in Electrical Engineering. Vol. 209,  pp. 111-119. (2013)  (doi: 10.1007/978-1-4471-4805-0_14)
18. Ahmad, I., Ayaz, S., Arafat S. Y., Riaz, F., Jabeen H. QoS routing for real time traffic in mobile ad hoc network // Proc. 7th Int. Conf. on Ubiquitous Information Management and Communication, ICUIMC '13 (2013) (doi: 10.1145/2448556.2448602)
19.  Liu R.D., Wu, W., Du J. D., Xin W., Yang D. C. QoS Oriented Traffic Modeling of Mobile POS Service in Cellular Network // Applied Mechanics and Materials, 2841, pp. 321-324, (2013) (doi: 10.4028/www.scientific.net/AMM.321-324.2841)
20. Abiteboul, S. Querying Semi-Structured Data // Proc. of 6th International Conference on Database Theory—ICDT'97, Delphi, pp. 1-18. (8-10 January 1997)
21. Luna J. M., Abdallah C. T. Control in computing systems: Part I // Computer-Aided Control System Design (CACSD), 2011 IEEE International Symposium, pp. 25-31. IEEE (2011) (doi: 10.1109/CACSD.2011.6044541)
22. Pluzhnik, E. V., Nikulchev E. V. Semistructured Database of Hybrid Cloud Computing Infrastructures // Modern Problems of Science and Education, No.3. (2013) http://www.science-education.ru/en/110-9980.